\documentclass[11pt,preprint]{aastex}
\usepackage{natbib}
\usepackage{color}
\usepackage{graphicx}
\usepackage{rotating}
\usepackage{hyperref}
\usepackage[all]{hypcap}
\usepackage{mathrsfs}
\usepackage{amsmath}
\usepackage{amssymb}
\hypersetup{
    bookmarks=true,         
    unicode=false,          
    pdftoolbar=true,        
    pdfmenubar=true,        
    pdffitwindow=false,     
    pdfstartview={FitH},    
    pdftitle={My title},    
    pdfauthor={Author},     
    pdfsubject={Subject},   
    pdfcreator={Creator},   
    pdfproducer={Producer}, 
    pdfkeywords={keyword1} {key2} {key3}, 
    pdfnewwindow=true,      
    colorlinks=true,        
    linkcolor=blue,         
    citecolor=blue,         
    filecolor=magenta,      
    urlcolor=cyan           
}

\begin{document}
\title{Solar Magnetic Flux Rope Eruption Simulated by a Data-Driven Magnetohydrodynamic Model}
\author{Yang Guo$^{1}$, Chun Xia$^{2}$, Rony Keppens$^{2,3}$, M. D. Ding$^1$, P. F. Chen$^1$}

\affil{$^1$ School of Astronomy and Space Science and Key Laboratory for Modern Astronomy and Astrophysics, Nanjing University, Nanjing 210023, China} \email{guoyang@nju.edu.cn}
\affil{$^2$ School of Physics and Astronomy, Yunnan University, Kunming 650050, China}
\affil{$^3$ Centre for mathematical Plasma Astrophysics, Department of Mathematics, KU Leuven, Celestijnenlaan 200B, B-3001 Leuven, Belgium}

\begin{abstract}
The combination of magnetohydrodynamic (MHD) simulation and multi-wavelength observations is an effective way to study mechanisms of magnetic flux rope eruption. We develop a data-driven MHD model using the zero-$\beta$ approximation. The initial condition is provided by nonlinear force-free field derived by the magneto-frictional method based on vector magnetic field observed by the Helioseismic and Magnetic Imager (HMI) aboard the \textit{Solar Dynamics Observatory} (\textit{SDO}). The bottom boundary uses observed time series of the vector magnetic field and the vector velocity derived by the Differential Affine Velocity Estimator for Vector Magnetograms (DAVE4VM). We apply the data-driven model to active region 11123 observed from 06:00 UT on 2011 November 11 to about 2 hours later. The evolution of the magnetic field topology coincides with the flare ribbons observed in the 304 and 1600 \AA \ wavebands by the Atmospheric Imaging Assembly. The morphology, propagation path, and propagation range of the flux rope are comparable with the observations in 304 \AA . We also find that a data-constrained boundary condition, where the bottom boundary is fixed to the initial values, reproduces a similar simulation result. This model can reproduce the evolution of a magnetic flux rope in its dynamic eruptive phase.

\end{abstract}


\section{Introduction}
Solar activity, such as prominence/filament eruptions, coronal mass ejections (CMEs) and flares are mostly related to magnetic flux rope eruptions, where the flux rope either pre-exists before or is formed during eruption \citep{2017Ouyang}. These eruptions expel magnetized plasma, with a huge amount of energy and helicity, and high energy particles into the interplanetary space and possibly disturb the space environment around the Earth. It has been proposed that one important driving mechanism for magnetic flux rope eruptions is due to ideal MHD instability, such as the torus instability (equivalent to the loss of equilibrium) for isolated flux ropes \citep{1988Demoulin,1991Forbes,2000Lin,2006Kliem,2010Demoulin,2014Kliem}, or tilt-kink instability for parallel current systems \citep{2014Keppens}, or coalescence-kink instability in anti-parallel current systems \citep{2018Makwana}. At the same time, the resistive process of magnetic reconnection could form the magnetic flux rope at the buildup phase, or facilitates the eruption in the driving phase \citep{1999Antiochos,2001Moore,2010Aulanier,2011Chen}. However, the interplay and feedback between ideal and resistive processes are still not fully understood. In this sense, it is crucial to build a working model for studying and, hopefully, predicting magnetic flux rope eruptions.

The buildup phase of a magnetic flux rope in a solar active region can be modeled with a sequence of magnetic equilibria, and has been widely studied by the nonlinear force-free field (NLFFF) models \citep[e.g.][]{2009Canou,2009Savcheva,2010Guo2,2012Inoue,2014Jiang2,2016Yang}. The dynamic process of magnetic flux rope eruptions can only be studied by magnetohydrodynamic (MHD) simulations. Here, we focus on a special class of such simulations, namely, data-driven and data-constrained MHD simulations, which are defined as simulations initiated and/or driven by observed data. Various data-driven numerical simulations differ mainly in three aspects. In the aspect of initial conditions, most of the data-driven and data-constrained MHD models are implemented with the initial condition provided by NLFFF models, derived by either the magneto-frictional \citep{2013Kliem,2016Jiang,2018Inoue} or the Grad-Rubin method \citep{2018Amari}. In the aspect of the boundary conditions, they could be provided from a time series of observations, which is called data-driven simulations \citep{2016Jiang}, or be assigned numerically with fixed values or zero-gradient extrapolations, which is called data-constrained simulations \citep{2013Kliem,2018Inoue,2018Amari}. In the aspect of the MHD model, either the zero-$\beta$ model \citep{2013Kliem,2018Inoue,2018Amari}, which omits the gas pressure, gravity, and the energy equation, or the ideal adiabatic MHD model \citep{2016Jiang} has been used thus far.

The data-driven and data-constrained MHD simulations have shown their strength in studying solar eruptions. For instance, we could study the buildup, triggering, and driving processes of a magnetic flux rope and its eruption; or investigate the roles of ideal and resistive processes in a flux rope eruption; or predict a flux rope eruption. Our motivation in this paper is to develop a data-driven MHD model and to directly confront it with observations. The observations of the vector magnetic field and extreme-ultraviolet (EUV) images are introduced in Section~\ref{sec:observation}. The MHD model is described in Section~\ref{sec:model}. The results are presented in Section~\ref{sec:result}. Summary and discussions are provided in Section~\ref{sec:summary}.

\section{Observations} \label{sec:observation}

We use the vector magnetic field, UV, and EUV observations in active region NOAA 11123 on 2010 November 11 as the input and benchmark for the data-driven MHD simulations. The vector magnetic field is obtained from the spectro-polarimetric observations provided by the Helioseismic and Magnetic Imager \citep[HMI;][]{2012Scherrer,2012Schou,2014Hoeksema} aboard the \textit{Solar Dynamics Observatory} \citep[\textit{SDO};][]{2012Pesnell}. Its cadence is 12 minutes and spatial sampling is $0.5''$ per pixel. The line-of-sight component of the vector magnetic field in active region 11123 at 06:00 UT is shown in Figure~\ref{fig:obser}a. Active region 11123 is a newly emerging region trailing behind the long-living preceding active region 11121. \citet{2014Mandrini} studied in detail the magnetic topology and eruptive activities in this active region complex. \citet{2015Galsgaard} studied the mechanisms of up-flows in this active region using a data-driven MHD simulation. In this study, we focus on the C4.7 class flare starting at 07:16 UT and peaking at 07:25 UT on 2010 November 11 accompanied with a small CME with a speed of about 250 km s$^{-1}$ \citep{2014Schmieder}. The Atmospheric Imaging Assembly \citep[AIA;][]{2012Lemen} aboard \textit{SDO} provides continuous observations in three UV-visible wavebands and seven EUV wavebands with a cadence of 12 seconds and spatial sampling of $0.6''$ per pixel. We use the UV and EUV images, especially in the 1600, 304, and 171~\AA \ wavebands, observed by \textit{SDO}/AIA to study the eruptive process and as the benchmark to test the MHD models. The pre-eruptive coronal loops and filament are shown in the 171~\AA \ and 304~\AA \ wavebands at 06:00 UT in Figure~\ref{fig:obser}c and \ref{fig:obser}d, respectively.

\section{MHD Model} \label{sec:model}

The lower corona is in a low-$\beta$ condition, where $\beta$ is the ratio between the gas pressure to the magnetic pressure. Thus, we adopt a zero-$\beta$ MHD model to simulate the dynamics in the corona. This model omits the gradient of gas pressure and gravity, and it also omits the energy equation. Only three physical variables, namely, the density, velocity, and the magnetic field, need to be solved:
\begin{equation}
\dfrac{\partial \rho}{\partial t} + \nabla \cdot (\rho \mathbf{v}) = \nu_\rho \nabla^2(\rho - \rho_\mathrm{i}), \label{eqn:mas}
\end{equation}
\begin{equation}
\dfrac{\partial (\rho \mathbf{v})}{\partial t} + \nabla \cdot (\mathbf{v} \rho \mathbf{v} - \mathbf{B}\mathbf{B}) + \nabla (\dfrac{\mathbf{B}^2}{2}) = \mu \nabla \cdot [2 \mathcal{S} - \dfrac{2}{3} (\nabla \cdot \mathbf{v}) \mathcal{I}], \label{eqn:mom}
\end{equation}
\begin{equation}
\dfrac{\partial \mathbf{B}}{\partial t} + \nabla \cdot (\mathbf{v} \mathbf{B} - \mathbf{B} \mathbf{v}) = -\nabla \times (\eta \mathbf{J}) , \label{eqn:ind}
\end{equation}
\begin{equation}
\mathbf{J} = \nabla \times \mathbf{B}, \label{eqn:cur}
\end{equation}
where $\rho$ is the density, $\mathbf{v}$ is the velocity, $\mathbf{B}$ is the magnetic field, $\nu_\rho$ is the density diffusion coefficient, $\rho_\mathrm{i}$ is the density at $t=0$, $\mu$ is the dynamic viscosity coefficient, $\mathcal{S}_{ij} = \frac{1}{2}(\partial v_i / \partial x_j + \partial v_j / \partial x_i)$ is the strain rate tensor, $\mathcal{I}$ is the unit tensor, and $\eta$ is the resistivity. Equations~(\ref{eqn:mas})--(\ref{eqn:cur}) are in the dimensionless form such that the vacuum permeability is assumed to be $\mu_0=1$. In the numerical setup, all the physical quantities and parameters are normalized by their corresponding typical factors, namely, $L_0 = 1.0 \times 10^9$ cm, $t_0 = 85.9$ s, $\rho_0 = 2.3 \times 10^{-15} \ \mathrm{g \ cm^{-3}}$, $\mathbf{v}_0$ = $1.2 \times 10^7 \ \mathrm{cm \ s^{-1}}$, and $\mathbf{B}_0 = 2.0$ G. The MHD equations are solved by the open source Message Passing Interface Adaptive Mesh Refinement Versatile Advection Code \citep[MPI-AMRVAC;][]{2003Keppens,2012Keppens,2014Porth,2018Xia}.

The source terms on the right hand side of Equations~(\ref{eqn:mas})--(\ref{eqn:ind}) are included with different purposes. Since the gradient of the gas pressure is omitted, the density evolution cannot be correctly handled without a density diffusion term in Equation~(\ref{eqn:mas}). This term is used to smooth the density distribution and maintain the initial density stratification in the evolution and the density diffusion coefficient is finally set to be $\nu_\rho=0.008$ by trial and error. The viscous term in Equation~(\ref{eqn:mom}) is adopted for numerical stability and the dynamic viscosity coefficient is set to be $\mu=0.05$. The resistive term in Equation~(\ref{eqn:ind}) is used to control the explicit magnetic diffusion and study the effect of magnetic reconnection. We use $\eta =0$ in the first two ideal cases and non-zero values in other cases.

The initial condition for the density $\rho$ is provided by a stratified atmosphere profile to simulate the solar atmosphere from the photosphere to the corona. We assume a stepwise function for the temperature:
\begin{equation}
T=
  \begin{cases}
    T_0                 & 0.0 \leq h < h_0  \\
    k_T (h-h_0) + T_0    & h_0 \leq h < h_1  \\
    T_1                 & h_1  \leq h < 13.8
  \end{cases},
\end{equation}
where $T_0=0.006$, $T_1=1.0$, $h_0=0.35$, $h_1=1.0$, and $k_T=(T_1 - T_0)/(h_1-h_0)$ in dimensionless units, or equivalently, $T_0=6.0 \times 10^3$~K, $T_1= 1.0 \times 10^6$~K, $h_0=3.5 \times 10^8$~cm, and $h_1= 1.0 \times 10^9$~cm in cgs units, which closely mimics the actual temperature-density variation with height. The density profile is derived by solving the hydrostatic equation:
\begin{equation}
\frac{\mathrm{d} p}{\mathrm{d} h} = -g \rho,
\end{equation}
where the gas pressure $p=\rho T$ in the dimensionless form. The density, $\rho$, on the bottom is chosen to be $1.0 \times 10^8$ (namely, $2.3 \times 10^{-7}$ g cm$^{-3}$), and it drops to 1.1 at $h=13.78$, i.e., the top of the computation box. Although we prescribe the density in this realistic fashion, the actual zero-$\beta$ simulation does not have pressure or temperature information, which is why we use the density diffusion in Equation~(\ref{eqn:mas}). The initial condition for the velocity is zero for all the three components, i.e., $\mathbf{v}=0$ at $t=0$.

\begin{figure}
\begin{center}
\includegraphics[width=1.0\textwidth]{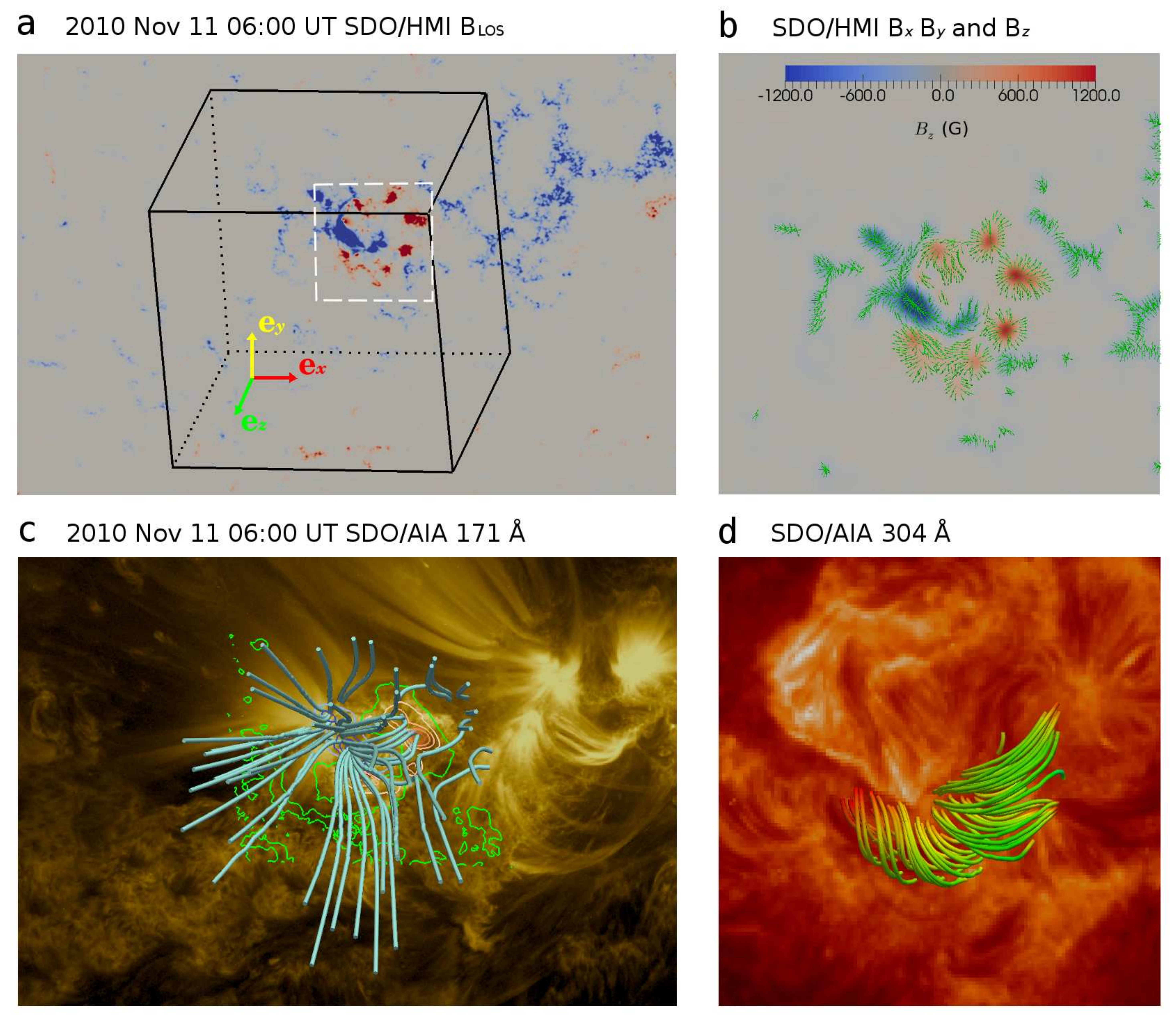}
\caption{(a) The background shows the line-of-sight magnetic field observed by \textit{SDO}/HMI at 06:00 UT on 2010 November 11 in active region 11123. The black box is the computational domain of the NLFFF and MHD models. The bottom of the black box and the white rectangle denote the field of views of panels (b) and (d), respectively. (b) The three components, $B_x$, $B_y$, and $B_z$ of the vector magnetic field, which has been projected to the heliographic coordinate system and preprocessed for the NLFFF modeling. (c) The background is the 171~\AA \ image observed by \textit{SDO}/AIA at 06:00 UT, whose field of view is the same as panel (a). The contours show the line-of-sight magnetic field at the bottom boundary of the computational domain, where the green contours indicate the polarity inversion line. Cyan solid curves represent magnetic field lines computed by the NLFFF model at 06:00 UT. (d) The background is the 304~\AA \ image observed by \textit{SDO}/AIA at 06:00 UT. Red-yellow-green solid curves represent magnetic field lines along the observed filament.
} \label{fig:obser}
\end{center}
\end{figure}

The initial condition for the magnetic field $\mathbf{B}$ is provided by the NLFFF model derived by the magneto-frictional method \citep{2016Guo1,2016Guo2}. We use the vector magnetic field by \textit{SDO}/HMI at 06:00 UT on 2010 November 11 as the boundary condition for constructing the NLFFF model. The series name for the vector magnetic field in the Joint Science Operations Center is ``hmi.B\_720s''. Two more processing steps in addition to the \textit{SDO}/HMI pipeline have been applied to the vector magnetic field. One is the correction of projection effects using the formula in \citet{1990Gary}. The other is preprocessing \citep{2006Wiegelmann} to remove the net Lorentz force and torque in order to conform to the force-free assumption. The projected and preprocessed vector magnetic field is shown in Figure~\ref{fig:obser}b. We use a vertically stretched grid to resolve the steep gradient of the density profile. The grid cell sizes from the bottom to top are geometric series. Therefore, the cell size in the vertical direction is $2.77 \times 10^{-3}$ on the bottom and 0.37 on the top, while the cell sizes on the horizontal directions are uniformly $0.76 \times 10^{-1}$. The computation box in the range $[x_\mathrm{min}, x_\mathrm{max}] \times [y_\mathrm{min}, y_\mathrm{max}] \times [z_\mathrm{min}, z_\mathrm{max}] = [-20.47,-6.79] \times [-35.57,-21.89] \times [0.1,13.78]$ is resolved by $180^3$ cells. The location and field of view of the computation box are shown in Figure~\ref{fig:obser}a. The finally derived NLFFF is displayed in Figure~\ref{fig:obser}c and \ref{fig:obser}d, which shows the comparisons between magnetic field lines with flare loops and a filament. We note that the computation box has been projected to the heliocentric coordinate system to compare with observations directly. The formulae to do the back projection are described in \cite{2017Guo}.

We need to specify the boundary conditions for density, velocity, and magnetic field  for each of the six boundaries ($x_\mathrm{min}, x_\mathrm{max}, y_\mathrm{min}, y_\mathrm{max}, z_\mathrm{min}, z_\mathrm{max}$). There are two ghost layers on each boundary to conform with the requirements of the numerical scheme, which is three-step time integration, HLL Riemann solver, and Koren limiter space reconstruction. To test the effects of the boundary conditions, we prepare two different cases:
\begin{description}
\item [Case I] data-driven boundary condition. The density is fixed to be the initial value on all the six boundaries. The velocity is fixed to be zero on the top and four side boundaries, and is set to be the data derived by the Differential Affine Velocity Estimator for Vector Magnetograms \citep[DAVE4VM;][]{2008Schuck} on the inner ghost layer of the bottom boundary, and zero-gradient extrapolation on the outer ghost layer of the bottom boundary. Since the velocity can only be derived at discrete times with a cadence of 12 minutes, the needed data with much higher cadence for the bottom boundary condition are computed with a linear interpolation in time. Finally, the magnetic field is provided by zero-gradient extrapolation on the top and four side boundaries, and is set to be the observed data on the inner ghost layer of the bottom boundary, and zero-gradient extrapolation on the outer ghost layer. Similar to the velocity, linear interpolation is used to fill the data gap between each magnetic field observation with a cadence of 12 minutes.

    We take some measures to guarantee the boundary condition conforming with the induction equation and $\nabla \cdot \mathbf{B}=0$ condition. First, we add a diffusive term $\delta \nabla (\nabla \cdot \mathbf{B})$ in the induction equation \citep{2003Keppens} to guarantee the solenoidal condition in the computation domain. Second, the normal component of the vector magnetic field on the outer ghost layer at the bottom boundary (and the two ghost layers at the other five boundaries) is reset by requiring $\nabla \cdot \mathbf{B}=0$ to guarantee the solenoidal condition for the boundaries. Third, the velocity field is derived by DAVE4VM, a method that is fully consistent with the induction equation \citep{2008Schuck}.

\item [Case II] data-constrained boundary condition. The velocity on the bottom boundary (both ghost layers) is fixed to be zero. The magnetic field on the inner ghost layer for the bottom boundary is fixed to be the initial observed data, namely, the vector magnetic field observed at 06:00 UT on 2010 November 11. The others are the same as Case I.
\end{description}

\section{Results} \label{sec:result}

\begin{figure}
\begin{center}
\includegraphics[width=0.8\textwidth]{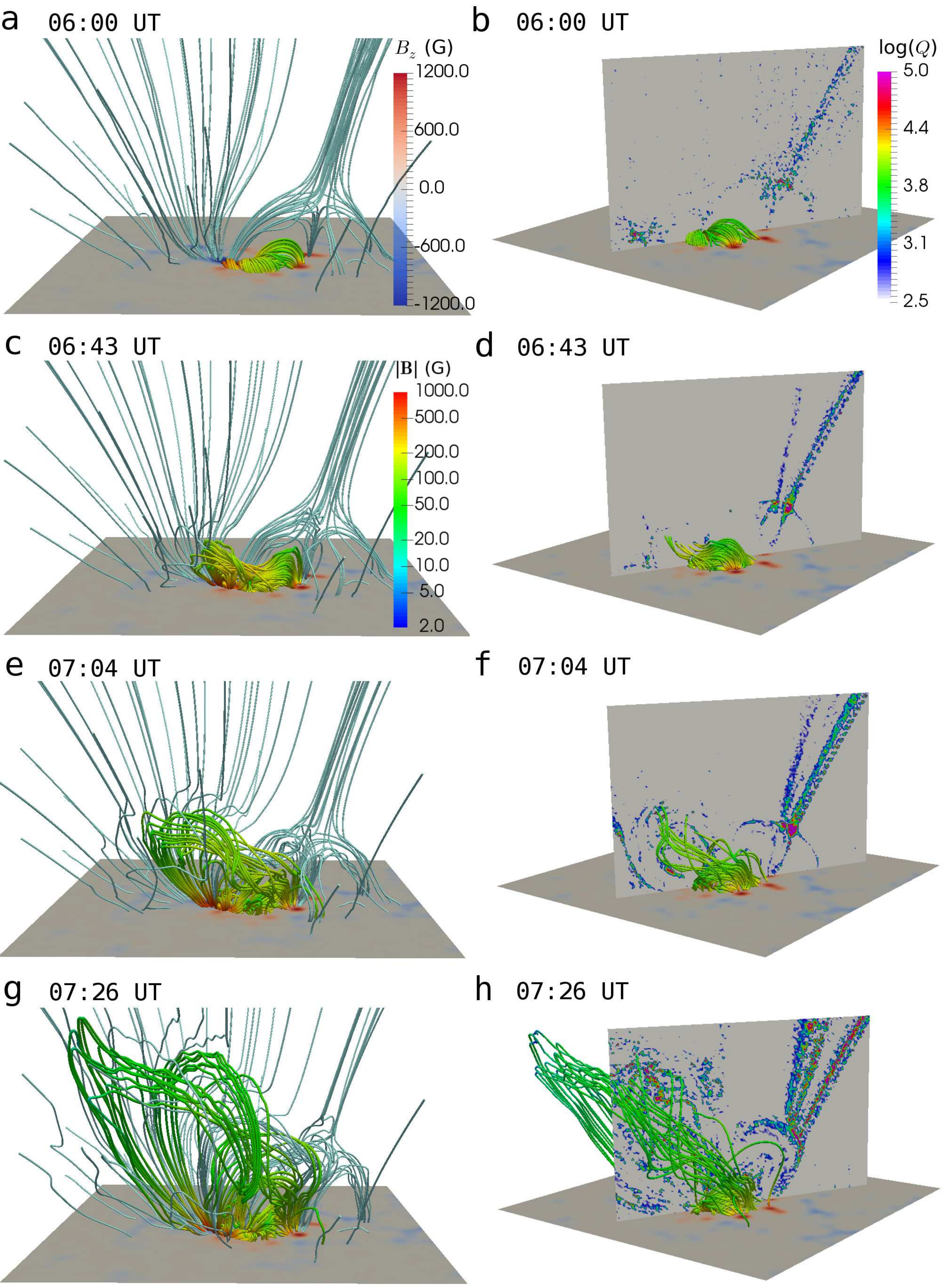}
\caption{Magnetic field lines and QSLs in the data-driven MHD simulations at four selected snapshots. (a) Magnetic field lines at 06:00 UT. The slice on the bottom shows $B_z$, whose color scale is indicated by the color bar. Cyan solid curves represent the background magnetic field lines. Red-yellow-green colored curves represent the sheared and twisted magnetic field lines. (b) QSLs at 06:00 UT. The color bar indicates the color scales for the logarithm of the squashing degree $Q$. (c, d) 06:43 UT. (e, f) 07:04 UT. (g, h) 07:26 UT. An animation is available in the online journal. The animation displays the evolution of the magnetic field lines from 06:00 UT to 07:41 UT. Panels (a), (c), (e), and (g) are four snapshots of the animation.
} \label{fig:mhd}
\end{center}
\end{figure}

The evolution of the magnetic field in the process of the flux rope eruption is shown in Figure~\ref{fig:mhd}. At 06:00 UT (Figure~\ref{fig:mhd}a), a magnetic flux rope as indicated by the sheared and twisted magnetic field lines lies under three magnetic null points \citep{2014Mandrini}, one of which is shown by the surrounding spine-fan shaped field lines. We also compute the quasi-separatrix layers (QSLs) to highlight the boundaries of different magnetic domains using a recently developed three-dimensional (3D) QSL computation method \citep{2012Pariat,2015Yang}. Figure~\ref{fig:mhd}b reveals that the flux rope is under the fan QSL. As time goes on, the two ends of the flux rope curve up (Figure~\ref{fig:mhd}c), and it stretches and distorts the fan QSL (Figure~\ref{fig:mhd}d). The flux rope is finally detached from the bottom and erupts into the corona as shown in Figure~\ref{fig:mhd}e in a sample snapshot at 07:04 UT. The QSLs surrounding and interleaving the flux rope also break the fan QSL (Figure~\ref{fig:mhd}f). At about 07:26 UT, the front of the flux rope rises to a high altitude and reaches the side boundary of the computation box (Figure~\ref{fig:mhd}g). The QSLs associated with it develop a very complex structure distributed over the boundary and body of the flux rope (Figure~\ref{fig:mhd}h). The propagation path of the front of the flux rope is not along the radial direction, but is inclined to the southeast.

\begin{figure}
\begin{center}
\includegraphics[width=1.0\textwidth]{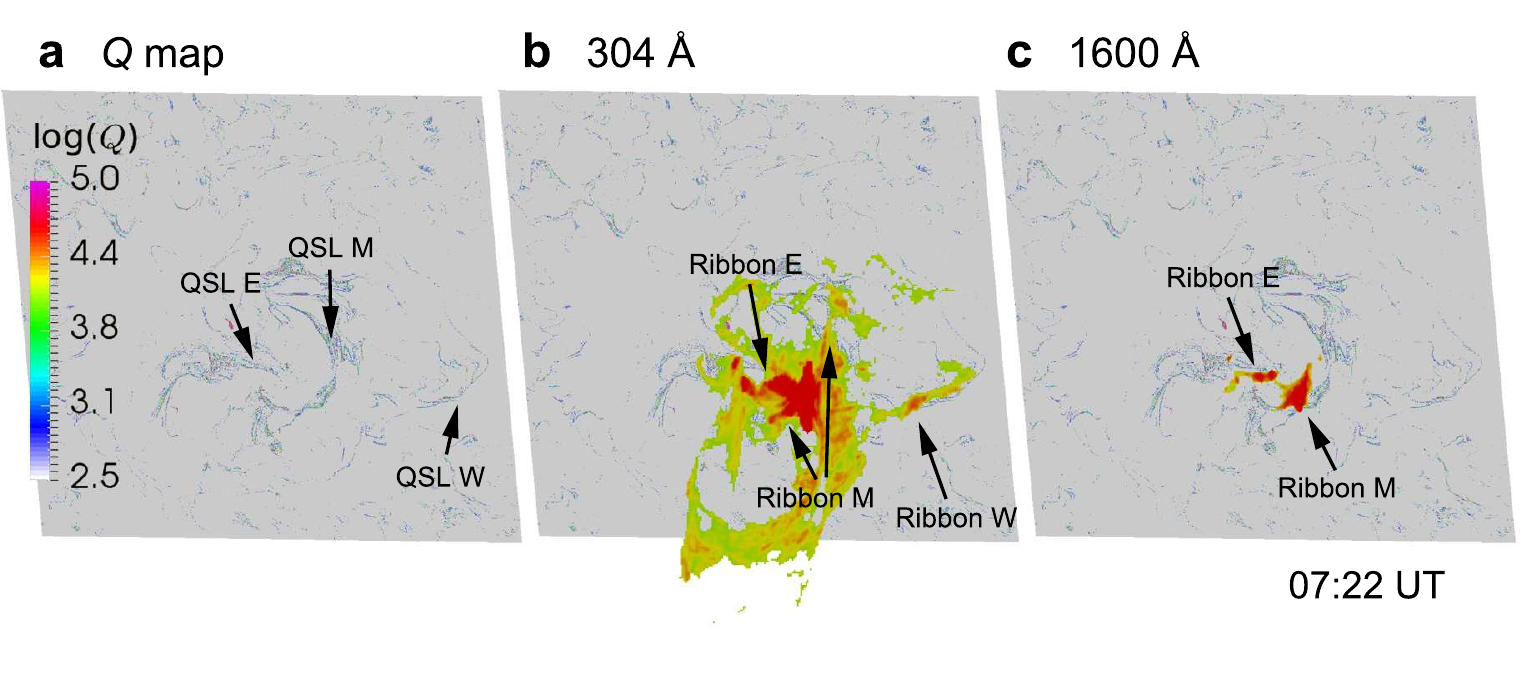}
\caption{(a) The squashing degree $Q$ at 07:22 UT on the bottom. (b) \textit{SDO}/AIA 304 \AA \ image at 07:22 UT overlaid on the $Q$ map. (c) \textit{SDO}/AIA 1600 \AA \ image at 07:22 UT overlaid on the $Q$ map.} \label{fig:qsl}
\end{center}
\end{figure}

Figure~\ref{fig:qsl} shows the comparison of the QSLs on the bottom with the flare ribbons observed by \textit{SDO}/AIA in both 304 and 1600~\AA \ wavebands. There are three major QSLs on the bottom as shown in Figure~\ref{fig:qsl}a at 07:22 UT, labeled the eastern QSL E, the middle QSL M, and the western QSL W, respectively. There are also three flare ribbons appearing in the 304~\AA \ image at 07:22 UT. It is found that Ribbons E, M, and W coincide spatially with parts of QSLs E, M, and W, respectively (Figure~\ref{fig:qsl}b). Two flare ribbons (E and M) appear in 1600~\AA , which also coincide with parts of the corresponding QSLs (Figure~\ref{fig:qsl}c). Flare ribbons always appear at the footpoints of QSLs. But QSLs could exist even in potential magnetic field with no electric current and free magnetic energy. Only when particles are accelerated or plasma are heated in QSLs, would flare ribbons appear at the footpoints of these QSLs. We note that when the QSLs are compared with the 304 and 1600~\AA \ observations, they have been back projected to heliocentric coordinates.

\begin{figure}
\begin{center}
\includegraphics[width=0.6\textwidth]{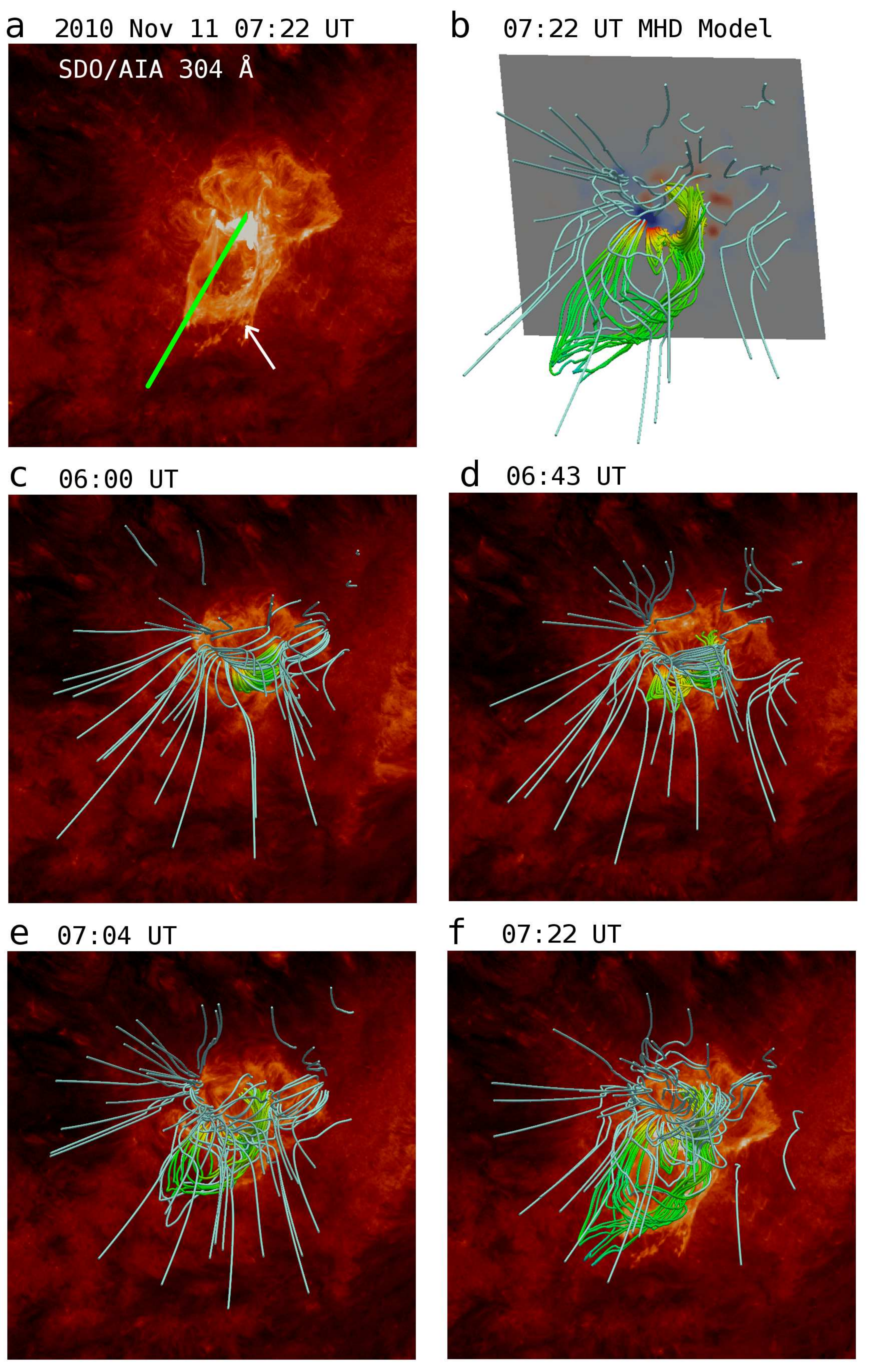}
\caption{(a) \textit{SDO}/AIA 304 \AA \ image at 07:22 UT on 2010 November 11. The green solid line shows the slice that we select to measure the time-distance profile of the erupting filament. The white arrow indicates the erupting filament material. (b) Magnetic field lines of the MHD model at 07:22 UT. (c) Magnetic field lines overlaid on the \textit{SDO}/AIA 304 \AA \ image at 06:00 UT. (d) 06:43 UT. (e) 07:04 UT. (f) 07:22 UT. An animation showing the time series of 304 \AA \ images, the evolution of magnetic field lines, and the magnetic field lines overlaid on the 304 \AA \ images from 06:00 UT to 07:41 UT is available in the online journal.} \label{fig:304mhd}
\end{center}
\end{figure}

The evolution of the flux rope simulated by the zero-$\beta$ MHD model is also compared with \textit{SDO}/AIA 304 \AA \ observations. The results in Figure~\ref{fig:304mhd} indicate that the MHD simulations conform with the observations in three aspects, namely, in morphology, eruption path, and propagation range. First, Figure~\ref{fig:304mhd}a shows a snapshot of the 304 \AA\ observation at 07:22 UT, where the erupting bright strand as indicated by the white arrow is generally believed to be a flux rope. The flux rope is highly curved, whose height is apparently larger than the distance between its two footpoints. Some flare loops under the erupting flux rope are so bright that their intensity is saturated in the 304~\AA \ image. The curved flux rope and the underlying cusp-shaped flare loops can also be found in the MHD model as shown in Figure~\ref{fig:304mhd}b. Second, the inclined eruption path is reproduced by the MHD model. It is not along the solar radial direction but inclined to the southeast. The projection of the eruption path coincides with the 304~\AA \ observations as shown in Figure \ref{fig:304mhd}c--\ref{fig:304mhd}f and the animation attached to Figure~\ref{fig:304mhd}. Finally, the flux rope in the observations erupted into interplanetary space and formed a CME \citep{2014Schmieder}. The animation attached to Figure~\ref{fig:304mhd} also indicates that the flux rope continues to erupt close to the boundary of the computation box.

\begin{figure}
\begin{center}
\includegraphics[width=1.0\textwidth]{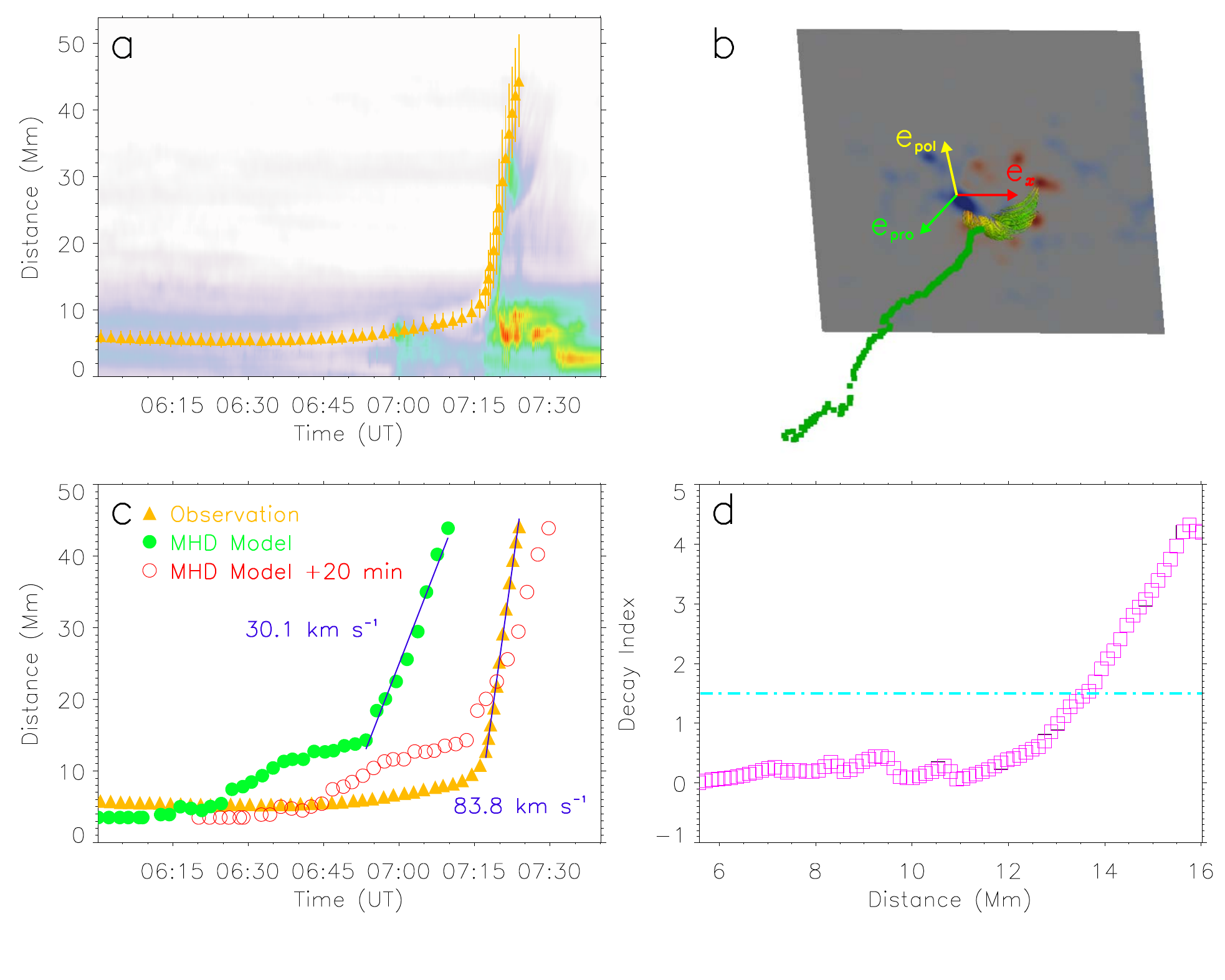}
\caption{(a) Time-distance measurement of the erupting flux rope. This image is the time stack of the image along the green line shown in Figure~\ref{fig:304mhd}a. Note that the lower-right direction in Figure~\ref{fig:304mhd}a is shown upward in this image. Orange triangles indicate the average position of ten measurements. Orange lines are the error bars. (b) The green dots show the front positions of the erupting flux rope in the MHD simulation. The arrows indicate three orthogonal directions defined in Section~\ref{sec:result}. (c) Time-distance profiles measured both in the 304~\AA \ observations and the MHD simulation. Green dots show the flux rope front in the original simulation, while red circles show that of the green ones being shifted 20 minutes later. The two solid lines are the linear fits to the rapid eruption stage of the flux rope. (d) The decay index distribution along a straight line, which is a nonlinear least-square regression of the 3D points in panel (b). } \label{fig:ht}
\end{center}
\end{figure}

Next, we compare quantitatively the eruption speeds of the flux rope both from the observation and from the MHD simulation. The speed observed in 304~\AA \ is measured by the time-distance diagram as shown in Figure~\ref{fig:ht}a, where the slice is indicated in Figure~\ref{fig:304mhd}a. We pinpoint the front of the flux rope at each observation time. The measurement is repeated ten times to estimate errors. We also measure the 3D positions of the flux rope front at selected snapshots in the MHD simulation (Figure~\ref{fig:ht}b). The projection of the 3D positions to the \textit{SDO} viewpoint is shown as the green dots in Figure~\ref{fig:ht}c. Similar to the time-distance profile in 304~\AA , the profile of the MHD simulation also shows a slow rising stage followed by a rapid rising stage. The only difference is that the flux rope starts to rise about 20 minutes earlier in the MHD simulation than that in the 304~\AA \ observation. Additionally, the flux rope rises a little bit faster and higher in the simulation than in the observation in the slow rising stage, while it is slower (30.1 km s$^{-1}$) in the simulation than in the observation (83.8 km s$^{-1}$) in the rapid rising stage. If we shift the time-distance profile in the simulation 20 minutes, the evolution processes are similar to each other as shown in Figure~\ref{fig:ht}c.

Finally, we compute the decay index along the eruption path of the flux rope to provide a qualitative explanation of the time-distance profile. Since the flux rope does not propagate along the radial direction of the Sun, the actual eruption direction is determined by a nonlinear least-square regression of the 3D points of the flux rope fronts, whose unit vector is defined as $\mathbf{e}_\mathrm{pro}$. The decay index is computed by $n = - \mathrm{d} \log B_\mathrm{ex,pol} / \mathrm{d} \log {s}$, where $B_\mathrm{ex,pol}$ is the field strength of the poloidal component of the external magnetic field, $s$ is the distance along the propagation path. Here, we use the potential field at the initial time, 06:00 UT, to represent $B_\mathrm{ex}$, whose poloidal component is defined as follows. We require that the unit vector of the poloidal direction, $\mathbf{e}_\mathrm{pol}$, is perpendicular to both $\mathbf{e}_\mathrm{pro}$ and the polarity inversion line, which is approximately along $\mathbf{e}_x$ in this case. These conditions require that $\mathbf{e}_\mathrm{pol} \times (\mathbf{e}_\mathrm{pro} \times \mathbf{e}_x) = 0$. The field strength of the poloidal component of the external magnetic field is computed by the projection of the potential field to the poloidal direction, $B_\mathrm{ex,pol}(s) = \mathbf{B}_\mathrm{pot}(s) \cdot \mathbf{e}_\mathrm{pol}$. The decay index at $s$ is thus obtained. It is not displayed against the distance $s$ along the propagation path, but is the projected distance in the plane-of-sky as shown in Figure~\ref{fig:ht}d to compare with the observations directly.

It is found that the flux rope starts to rise rapidly at about 14.0 Mm as shown by the green dots in Figure~\ref{fig:ht}c. And at about 13.7 Mm (Figure~\ref{fig:ht}d) the decay index crosses the canonical critical value of 1.5. We note that the critical decay index of torus instability is not strictly 1.5 but ranges from 1 to 2 depending on different current paths (straight, semicircular or others), which has been found by either theoretical or numerical methods \citep{1978Bateman,1978vanTend,2006Kliem,2007Torok,2010Demoulin,2010Fan,2010Olmedo}. Taking the range of 1 to 2 and referring to Figure~\ref{fig:ht}d, it is found that the torus unstable distance ranges from about 13.0 to 14.0 Mm, which is a small range and is consistent with the turning position (14.0 Mm) where the eruption velocity changes from slow to fast. This result is in agreement with the prediction of the torus instability. From Figure~\ref{fig:ht}c, we find that the flux rope starts to rise at about 5.0 Mm as shown by the green dots. This distance is far below the torus unstable region as found in Figure~\ref{fig:ht}d using the decay index range of 1 to 2. Therefore, the flux rope needs additional physical mechanisms to rise to the torus unstable region. Magnetic reconnection in null points or below the flux rope (similar to the tether-cutting reconnection) could perform such a role to raise a flux rope to the torus unstable region.

\section{Summary and Discussion} \label{sec:summary}

We have developed a data-driven MHD model using the zero-$\beta$ MHD equations, which are solved by the open-source code MPI-AMRVAC \citep{{2003Keppens,2012Keppens,2014Porth,2018Xia}}. This model is applied to active region 11123 that was observed by \textit{SDO}/HMI and \textit{SDO}/AIA. The initial condition of the MHD simulation is provided by the NLFFF model constructed from the vector magnetic field at 06:00 UT on 2010 November 11 and the magneto-frictional method \citep{2016Guo1,2016Guo2}. The boundary condition on the bottom is based on vector magnetic field observations and vector velocity derived by the DAVE4VM method \citep{2008Schuck}. The data-driven MHD simulation reproduces the eruption process of the magnetic flux rope. The QSLs in the vertical direction show that the flux rope lies under a complex spine-fan structure, which contains three magnetic null points \citep{2014Mandrini}. Parts of the QSLs on the bottom surface coincide with the flare ribbons observed in 304 and 1600 \AA . The morphology, propagation path, and propagation range of the flux rope eruption in the MHD simulation resemble those in the \textit{SDO}/AIA 304~\AA \ observations. The height-time profiles of the flux rope front in both the MHD simulation and the 304~\AA \ observations have a two stage evolution, one of which is slow and the other is rapid. These similarities demonstrate physics-based numerical predictions of solar eruptions. However, it is noted that the flux rope in the MHD simulation erupts about 20 minutes earlier than the observation, and the velocity of the flux rope is slower in the simulation than that in the observation. These discrepancies indicate the limitations of the present zero-$\beta$ simulation, which asks for further improvements, such as including the energy equation and considering the effects of thermal conduction and radiative cooling.

To test the effects of different boundary conditions, we prepare a data-constrained boundary, namely Case II as described in Section~\ref{sec:model}. The simulation results of Case II are very similar to Case I in terms of morphology, propagation path and range of the flux rope. This similarity reveals two noteworthy aspects of the MHD model. First, magnetic flux rope eruptions possess varying temporal and spatial scales. Compared to the dynamic eruptive phase, the buildup phase of a magnetic flux rope is relatively long, which may last for several days or even weeks. The data-driven effects should be more important in this long phase than in the eruptive one. Therefore, the data-driven boundary condition is indispensable in modeling the buildup phase, while it might be negligible close to an eruption. Second, this result implies that we could predict the eruption of a magnetic flux rope in principle. Since the effect of the boundary condition could be neglected when initial conditions are close enough to the dynamic eruptive phase, we might only use data in advance of the eruption to predict this following evolution. Furthermore, the apparent indifference of our simulation between the data-driven and data-constrained approaches confirms once more that ideal MHD instabilities, such as the torus instability in this particular configuration, unavoidably lead to eruptions. The initial rise to build up the torus unstable state in the data-constrained simulation might be caused by the unbalanced Lorentz force in the numerical NLFFF model, or a finite but small resistivity in the numerical diffusion of the simulation. Since it is well-known that subtle details in internal pitch or external field variations play a decisive role in determining the growth rate of ideal MHD instabilities \citep[see, e.g.,][]{2004Goedbloed}, it remains challenging to get precise agreements in full eruption dynamics.

We note that the above two cases are only influenced by numerical resistivity since the explicit resistivity was set to be 0 so far. We also investigate the effects of tuning the resistivity, $\eta$, in Equation~(\ref{eqn:ind}). Two sets of tests are performed. In the first set, $\eta$ is uniform in the whole computation box. It is found that when $\eta \le 5 \times 10^{-5}$, the simulation result is very similar to that with $\eta = 0$. When $\eta \ge 2 \times 10^{-4}$, we find noticeable differences in the simulation result, where the rising speed of the erupting flux rope is slower and the regions with large electric current start to change drastically. The observations do not show such changes. The experiments imply that the resistivity, which is derived from our inherent numerical diffusion in the previous $\eta=0$ runs, is of an acceptable level to obtain CME eruptions that are comparable with observations. In the second set, $\eta=\eta_0$ when $J < J_c$, and $\eta = \eta_2 [(J-J_c)/J_c]^2 + \eta_0$ when $J \ge J_c$, where $\eta_0$ and $\eta_2$ are the resistivity of the background and the coefficient of anomalous resistivity, respectively, and $J_c$ is the critical electric current density. In a reasonable range of these parameters, the results do not show better evolution of the magnetic field compared to observations. A possible reason is that the excitation of anomalous resistivity should not be determined by the current density only. To simulate the effect of magnetic reconnection in accordance with observations, we need to specify its location and timing with more constraints or resort to more advanced coupled MHD and particle in cell (PIC) methods, as recently demonstrated on coalescing islands in \citet{2018Makwana}.


\acknowledgments

The authors thank the anonymous referee for constructive comments. The \textit{SDO} data are available by courtesy of NASA/\textit{SDO} and the AIA and HMI science teams. YG thanks Chaowei Jiang for useful discussions on the boundary condition. YG, MDD, and PFC are supported by NSFC (11773016, 11733003, and 11533005), NKBRSF 2014CB744203, the fundamental research funds for the central universities 020114380028, and Jiangsu 333 Project (BRA2017359). CX is supported by NSFC (11803031). RK is supported by KU Leuven (GOA/2015-014). The numerical calculations in this paper have been done using the computing facilities of the High Performance Computing Center (HPCC) in Nanjing University.



\begin{thebibliography}{49}
\expandafter\ifx\csname natexlab\endcsname\relax\def\natexlab#1{#1}\fi

\bibitem[{{Amari} {et~al.}(2018){Amari}, {Canou}, {Aly}, {Delyon}, \&
  {Alauzet}}]{2018Amari}
{Amari}, T., {Canou}, A., {Aly}, J.-J., {Delyon}, F., \& {Alauzet}, F. 2018,
  \nat, 554, 211

\bibitem[{{Antiochos} {et~al.}(1999){Antiochos}, {DeVore}, \&
  {Klimchuk}}]{1999Antiochos}
{Antiochos}, S.~K., {DeVore}, C.~R., \& {Klimchuk}, J.~A. 1999, \apj, 510, 485

\bibitem[{{Aulanier} {et~al.}(2010){Aulanier}, {T{\"o}r{\"o}k}, {D{\'e}moulin},
  \& {DeLuca}}]{2010Aulanier}
{Aulanier}, G., {T{\"o}r{\"o}k}, T., {D{\'e}moulin}, P., \& {DeLuca}, E.~E.
  2010, \apj, 708, 314

\bibitem[{{Bateman}(1978)}]{1978Bateman}
{Bateman}, G. 1978, {MHD Instabilities} (Cambridge, MA: MIT Press)

\bibitem[{{Canou} {et~al.}(2009){Canou}, {Amari}, {Bommier}, {Schmieder},
  {Aulanier}, \& {Li}}]{2009Canou}
{Canou}, A., {Amari}, T., {Bommier}, V., {Schmieder}, B., {Aulanier}, G., \&
  {Li}, H. 2009, \apjl, 693, L27

\bibitem[{{Chen}(2011)}]{2011Chen}
{Chen}, P.~F. 2011, Living Reviews in Solar Physics, 8, 1

\bibitem[{{D{\'e}moulin} \& {Aulanier}(2010)}]{2010Demoulin}
{D{\'e}moulin}, P. \& {Aulanier}, G. 2010, \apj, 718, 1388

\bibitem[{{D\'{e}moulin} \& {Priest}(1988)}]{1988Demoulin}
{D\'{e}moulin}, P. \& {Priest}, E.~R. 1988, \aap, 206, 336

\bibitem[{{Fan}(2010)}]{2010Fan}
{Fan}, Y. 2010, \apj, 719, 728

\bibitem[{{Forbes} \& {Isenberg}(1991)}]{1991Forbes}
{Forbes}, T.~G. \& {Isenberg}, P.~A. 1991, \apj, 373, 294

\bibitem[{{Galsgaard} {et~al.}(2015){Galsgaard}, {Madjarska}, {Vanninathan},
  {Huang}, \& {Presmann}}]{2015Galsgaard}
{Galsgaard}, K., {Madjarska}, M.~S., {Vanninathan}, K., {Huang}, Z., \&
  {Presmann}, M. 2015, \aap, 584, A39

\bibitem[{{Gary} \& {Hagyard}(1990)}]{1990Gary}
{Gary}, G.~A. \& {Hagyard}, M.~J. 1990, \solphys, 126, 21

\bibitem[{{Goedbloed} \& {Poedts}(2004)}]{2004Goedbloed}
{Goedbloed}, J.~P.~H. \& {Poedts}, S. 2004, {Principles of
  Magnetohydrodynamics}

\bibitem[{{Guo} {et~al.}(2017){Guo}, {Cheng}, \& {Ding}}]{2017Guo}
{Guo}, Y., {Cheng}, X., \& {Ding}, M. 2017, Science in China Earth Sciences,
  60, 1408

\bibitem[{{Guo} {et~al.}(2010){Guo}, {Schmieder}, {D{\'e}moulin}, {Wiegelmann},
  {Aulanier}, {T{\"o}r{\"o}k}, \& {Bommier}}]{2010Guo2}
{Guo}, Y., {Schmieder}, B., {D{\'e}moulin}, P., {Wiegelmann}, T., {Aulanier},
  G., {T{\"o}r{\"o}k}, T., \& {Bommier}, V. 2010, \apj, 714, 343

\bibitem[{{Guo} {et~al.}(2016{\natexlab{a}}){Guo}, {Xia}, {Keppens}, \&
  {Valori}}]{2016Guo1}
{Guo}, Y., {Xia}, C., {Keppens}, R., \& {Valori}, G. 2016{\natexlab{a}}, \apj,
  828, 82

\bibitem[{{Guo} {et~al.}(2016{\natexlab{b}}){Guo}, {Xia}, \&
  {Keppens}}]{2016Guo2}
{Guo}, Y., {Xia}, C., \& {Keppens}, R. 2016{\natexlab{b}}, \apj, 828, 83

\bibitem[{{Hoeksema} {et~al.}(2014){Hoeksema}, {Liu}, {Hayashi}, {Sun},
  {Schou}, {Couvidat}, {Norton}, {Bobra}, {Centeno}, {Leka}, {Barnes}, \&
  {Turmon}}]{2014Hoeksema}
{Hoeksema}, J.~T., {Liu}, Y., {Hayashi}, K., {Sun}, X., {Schou}, J.,
  {Couvidat}, S., {Norton}, A., {Bobra}, M., {Centeno}, R., {Leka}, K.~D.,
  {Barnes}, G., \& {Turmon}, M. 2014, \solphys, 289, 3483

\bibitem[{{Inoue} {et~al.}(2018){Inoue}, {Kusano}, {B{\"u}chner}, \&
  {Sk{\'a}la}}]{2018Inoue}
{Inoue}, S., {Kusano}, K., {B{\"u}chner}, J., \& {Sk{\'a}la}, J. 2018, Nature
  Communications, 9, 174

\bibitem[{{Inoue} {et~al.}(2012){Inoue}, {Shiota}, {Yamamoto}, {Pandey},
  {Magara}, \& {Choe}}]{2012Inoue}
{Inoue}, S., {Shiota}, D., {Yamamoto}, T.~T., {Pandey}, V.~S., {Magara}, T., \&
  {Choe}, G.~S. 2012, \apj, 760, 17

\bibitem[{{Jiang} {et~al.}(2014){Jiang}, {Wu}, {Feng}, \& {Hu}}]{2014Jiang2}
{Jiang}, C., {Wu}, S.~T., {Feng}, X., \& {Hu}, Q. 2014, \apjl, 786, L16

\bibitem[{{Jiang} {et~al.}(2016){Jiang}, {Wu}, {Feng}, \& {Hu}}]{2016Jiang}
---. 2016, Nature Communications, 7, 11522

\bibitem[{{Keppens} {et~al.}(2012){Keppens}, {Meliani}, {van Marle}, {Delmont},
  {Vlasis}, \& {van der Holst}}]{2012Keppens}
{Keppens}, R., {Meliani}, Z., {van Marle}, A.~J., {Delmont}, P., {Vlasis}, A.,
  \& {van der Holst}, B. 2012, Journal of Computational Physics, 231, 718

\bibitem[{{Keppens} {et~al.}(2003){Keppens}, {Nool}, {T{\'o}th}, \&
  {Goedbloed}}]{2003Keppens}
{Keppens}, R., {Nool}, M., {T{\'o}th}, G., \& {Goedbloed}, J.~P. 2003, Computer
  Physics Communications, 153, 317

\bibitem[{{Keppens} {et~al.}(2014){Keppens}, {Porth}, \& {Xia}}]{2014Keppens}
{Keppens}, R., {Porth}, O., \& {Xia}, C. 2014, \apj, 795, 77

\bibitem[{{Kliem} {et~al.}(2014){Kliem}, {Lin}, {Forbes}, {Priest}, \&
  {T{\"o}r{\"o}k}}]{2014Kliem}
{Kliem}, B., {Lin}, J., {Forbes}, T.~G., {Priest}, E.~R., \& {T{\"o}r{\"o}k},
  T. 2014, \apj, 789, 46

\bibitem[{{Kliem} {et~al.}(2013){Kliem}, {Su}, {van Ballegooijen}, \&
  {DeLuca}}]{2013Kliem}
{Kliem}, B., {Su}, Y.~N., {van Ballegooijen}, A.~A., \& {DeLuca}, E.~E. 2013,
  \apj, 779, 129

\bibitem[{{Kliem} \& {T{\"o}r{\"o}k}(2006)}]{2006Kliem}
{Kliem}, B. \& {T{\"o}r{\"o}k}, T. 2006, Physical Review Letters, 96, 255002

\bibitem[{{Lemen} {et~al.}(2012){Lemen}, {Title}, {Akin}, {Boerner}, {Chou},
  {Drake}, {et~al.}}]{2012Lemen}
{Lemen}, J.~R., {Title}, A.~M., {Akin}, D.~J., {Boerner}, P.~F., {Chou}, C.,
  {Drake}, J.~F., {et~al.} 2012, \solphys, 275, 17

\bibitem[{{Lin} \& {Forbes}(2000)}]{2000Lin}
{Lin}, J. \& {Forbes}, T.~G. 2000, \jgr, 105, 2375

\bibitem[{{Makwana} {et~al.}(2018){Makwana}, {Keppens}, \&
  {Lapenta}}]{2018Makwana}
{Makwana}, K.~D., {Keppens}, R., \& {Lapenta}, G. 2018, Physics of Plasmas, 25,
  082904

\bibitem[{{Mandrini} {et~al.}(2014){Mandrini}, {Schmieder}, {D{\'e}moulin},
  {Guo}, \& {Cristiani}}]{2014Mandrini}
{Mandrini}, C.~H., {Schmieder}, B., {D{\'e}moulin}, P., {Guo}, Y., \&
  {Cristiani}, G.~D. 2014, \solphys, 289, 2041

\bibitem[{{Moore} {et~al.}(2001){Moore}, {Sterling}, {Hudson}, \&
  {Lemen}}]{2001Moore}
{Moore}, R.~L., {Sterling}, A.~C., {Hudson}, H.~S., \& {Lemen}, J.~R. 2001,
  \apj, 552, 833

\bibitem[{{Olmedo} \& {Zhang}(2010)}]{2010Olmedo}
{Olmedo}, O. \& {Zhang}, J. 2010, \apj, 718, 433

\bibitem[{{Ouyang} {et~al.}(2017){Ouyang}, {Zhou}, {Chen}, \&
  {Fang}}]{2017Ouyang}
{Ouyang}, Y., {Zhou}, Y.~H., {Chen}, P.~F., \& {Fang}, C. 2017, \apj, 835, 94

\bibitem[{{Pariat} \& {D{\'e}moulin}(2012)}]{2012Pariat}
{Pariat}, E. \& {D{\'e}moulin}, P. 2012, \aap, 541, A78

\bibitem[{{Pesnell} {et~al.}(2012){Pesnell}, {Thompson}, \&
  {Chamberlin}}]{2012Pesnell}
{Pesnell}, W.~D., {Thompson}, B.~J., \& {Chamberlin}, P.~C. 2012, \solphys,
  275, 3

\bibitem[{{Porth} {et~al.}(2014){Porth}, {Xia}, {Hendrix}, {Moschou}, \&
  {Keppens}}]{2014Porth}
{Porth}, O., {Xia}, C., {Hendrix}, T., {Moschou}, S.~P., \& {Keppens}, R. 2014,
  \apjs, 214, 4

\bibitem[{{Savcheva} \& {van Ballegooijen}(2009)}]{2009Savcheva}
{Savcheva}, A. \& {van Ballegooijen}, A. 2009, \apj, 703, 1766

\bibitem[{{Scherrer} {et~al.}(2012){Scherrer}, {Schou}, {Bush}, {Kosovichev},
  {Bogart}, {Hoeksema}, {et~al.}}]{2012Scherrer}
{Scherrer}, P.~H., {Schou}, J., {Bush}, R.~I., {Kosovichev}, A.~G., {Bogart},
  R.~S., {Hoeksema}, J.~T., {et~al.} 2012, \solphys, 275, 207

\bibitem[{{Schmieder} {et~al.}(2014){Schmieder}, {Cremades}, {Mandrini},
  {D{\'e}moulin}, \& {Guo}}]{2014Schmieder}
{Schmieder}, B., {Cremades}, H., {Mandrini}, C., {D{\'e}moulin}, P., \& {Guo},
  Y. 2014, in IAU Symposium, Vol. 300, Nature of Prominences and their Role in
  Space Weather, ed. B.~{Schmieder}, J.-M. {Malherbe}, \& S.~T. {Wu}, 489--490

\bibitem[{{Schou} {et~al.}(2012){Schou}, {Scherrer}, {Bush}, {Wachter},
  {Couvidat}, {Rabello-Soares}, {et~al.}}]{2012Schou}
{Schou}, J., {Scherrer}, P.~H., {Bush}, R.~I., {Wachter}, R., {Couvidat}, S.,
  {Rabello-Soares}, M.~C., {et~al.} 2012, \solphys, 275, 229

\bibitem[{{Schuck}(2008)}]{2008Schuck}
{Schuck}, P.~W. 2008, \apj, 683, 1134

\bibitem[{{T{\"o}r{\"o}k} \& {Kliem}(2007)}]{2007Torok}
{T{\"o}r{\"o}k}, T. \& {Kliem}, B. 2007, Astronomische Nachrichten, 328, 743

\bibitem[{{van Tend} \& {Kuperus}(1978)}]{1978vanTend}
{van Tend}, W. \& {Kuperus}, M. 1978, \solphys, 59, 115

\bibitem[{{Wiegelmann} {et~al.}(2006){Wiegelmann}, {Inhester}, \&
  {Sakurai}}]{2006Wiegelmann}
{Wiegelmann}, T., {Inhester}, B., \& {Sakurai}, T. 2006, \solphys, 233, 215

\bibitem[{{Xia} {et~al.}(2018){Xia}, {Teunissen}, {El Mellah}, {Chan{\'e}}, \&
  {Keppens}}]{2018Xia}
{Xia}, C., {Teunissen}, J., {El Mellah}, I., {Chan{\'e}}, E., \& {Keppens}, R.
  2018, \apjs, 234, 30

\bibitem[{{Yang} {et~al.}(2015){Yang}, {Guo}, \& {Ding}}]{2015Yang}
{Yang}, K., {Guo}, Y., \& {Ding}, M.~D. 2015, \apj, 806, 171

\bibitem[{{Yang} {et~al.}(2016){Yang}, {Guo}, \& {Ding}}]{2016Yang}
---. 2016, \apj, 824, 148

\end{thebibliography}



\end{document}